\documentclass[graybox]{svmult}

\usepackage{mathptmx}       
\usepackage{helvet}         
\usepackage{courier}        
\usepackage{type1cm}        
\usepackage{makeidx}         
\usepackage{graphicx}        
\usepackage{multicol}        
\usepackage[bottom]{footmisc}

\makeindex             

\begin{document}

\title*{Black holes in extended gravity theories in Palatini formalism}
\author{Jesus Martinez-Asencio, Gonzalo J. Olmo and D. Rubiera-Garcia}
\institute{G. J. Olmo and J. Martinez-Asencio \at Departamento de F\'{i}sica Te\'{o}rica and
IFIC,  Universidad de Valencia-CSIC, Facultad de F\'{i}sica, C/ Dr. Moliner 50,
Burjassot-46100, Valencia, Spain.\email{gonzalo.olmo@csic.es}
\and D. Rubiera-Garcia \at Departamento de F\'isica, Universidad de Oviedo. Avenida
Calvo Sotelo 18, 33007 Oviedo, Asturias, Spain. \email{rubieradiego@gmail.com}}

%
%
\maketitle

\abstract{We consider several physical scenarios where black holes within classical gravity theories including $R^2$ and Ricci-squared corrections and formulated \`a la Palatini can be analytically studied.}

\section{Palatini approach and black holes}
\label{sec:1}

In General Relativity (GR) the field equations are obtained by assuming a Riemannian geometry and performing variations of the action with respect to the metric. If the Riemannian condition is relaxed and the connection is allowed to vary independently of the metric (Palatini formalism) the resulting field equations for GR turn out to be equivalent to those obtained under the Riemannian assumption  though, in general, they are different. This is so because in the case of GR the connection field equation can be solved in terms of the Christoffel symbols of the metric (Levi-Civita connection). This compatibility between metric and connection and thus the identification between both approaches has mostly become  implicitly assumed in any extension of GR that includes curvature  corrections  to the Einstein-Hilbert action. We note that, in particular, quadratic curvature terms are required for a high-energy completion of the theory and indeed arise in the quantization of fields in curved spacetime \cite{quantization} and in several approaches to quantum gravity such as those based on string theory \cite{string}. However, the physical and mathematical properties of extended gravity theories formulated in the metric or Palatini approaches are very different. While the former is affected by a number of problems such as higher-order field equations or the existence of ghosts and other perturbative instabilities, the latter provides always second-order field equations and absence of ghosts. In the particular case of Lovelock gravities \cite{Lovelock}, metric and Palatini formulations yield the same field equations \cite{Borunda}. Here we briefly summarize some scenarios concerning black holes in quadratic (Palatini) extensions of GR, where the connection equation can be algebraically solved and the properties of the theories formulated in this way are physically appealing.

For a theory $f(R,Q)$ depending on the invariants $R=g_{\mu\nu}R^{\mu\nu}$ and $Q=R_{\mu\nu}R^{\mu\nu}$ the variation with respect to the connection $\Gamma_{\alpha\beta}^{\gamma}$ leads, in the most general case (admitting both nonvanishing torsion $\Gamma_{[\alpha \beta]}^{\gamma}$ and antisymmetric-Ricci $R_{[\mu\nu]}$), to \cite{torsion}
\begin{eqnarray}
\frac{1}{\sqrt{-g}}\widetilde{\nabla}_{\alpha}[\sqrt{-g}M^{(\beta \nu)}]&=&\Xi_{\alpha}^{(+) \beta \nu \kappa \rho} M_{[\kappa \rho]} \label{eq:con1a} \\
\frac{1}{\sqrt{-g}}\widetilde{\nabla}_{\alpha}[\sqrt{-g}M^{[\beta \nu]}]&=&\Xi_{\alpha}^{(-) \beta \nu \kappa \rho} M_{(\kappa \rho)} \label{eq:con1b}
\end{eqnarray}
where $M^{(\beta \nu)}=f_R g^{\beta \nu} + 2f_{Q} R^{(\beta \nu)}(\Gamma)$, $M^{[\beta \nu]}=2f_{Q} R^{[\beta \nu]}(\Gamma)$ and $\Xi_{\alpha}^{(\pm) \beta \nu \kappa \rho}=[\widetilde{S}_{\alpha \lambda}^{\nu} g^{\beta \kappa} \pm \widetilde{S}_{\alpha \lambda}^{\beta} g^{\nu \kappa}] g^{\lambda \rho}$. The symmetric connection $\widetilde{C}_{\alpha \lambda}^{\beta}$ in the covariant derivative $\nabla^{\widetilde{C}}_{\alpha}$ in (\ref{eq:con1a}) and (\ref{eq:con1b}) has been introduced by convenience, and is related to the original connection $\Gamma_{\alpha \lambda}^{\beta}=C_{\alpha \lambda}^{\beta}+S_{\alpha \lambda}^{\beta}$ (with $C_{\alpha \lambda}^{\beta}$ and $S_{\alpha \lambda}^{\beta}$ its symmetric and antisymmetric components, respectively) as (with $S_{\mu}=S_{\mu\lambda}^{\lambda}$)
\begin{equation}
C_{\mu\nu}^{\lambda}=\widetilde{C}_{\mu\nu}^{\lambda}-\frac{1}{3} (\delta_{\nu}^{\lambda}S_{\mu}+\delta_{\mu}^{\lambda} S_{\nu}) \hspace{0.1cm}; \hspace{0.1cm}
S_{\mu\nu}^{\lambda}=\widetilde{S}_{\mu\nu}^{\lambda}-\frac{1}{3} (\delta_{\nu}^{\lambda}S_{\mu}-\delta_{\mu}^{\lambda} S_{\nu}).
\end{equation}

Equations (\ref{eq:con1a}) and (\ref{eq:con1b}) constitute a system such that the symmetric and antisymmetric parts of $M^{\mu\nu}$ are coupled to each other through the torsion tensor $\widetilde{S}_{\alpha \lambda}^{\beta}$. The simplest solutions to this system of equations correspond to putting both $\widetilde{S}_{\alpha \lambda}^{\beta}$ and $R_{[\mu\nu]}$ to zero, so that (\ref{eq:con1b}) becomes trivially satisfied and (\ref{eq:con1a}) reads simply

\begin{equation}
\frac{1}{\sqrt{-g}}\nabla_{\alpha}[\sqrt{-g}(f_R g^{\mu\nu} + 2f_Q R^{(\beta \nu)})]=0 \label{eq:con3}
\end{equation}
On the other hand, the variation of the action with respect to metric leads to

\begin{equation}
f_R R_{\mu\nu}-\frac{f}{2}g_{\mu\nu}+2f_Q R_{\mu\alpha}{R^\alpha}_\nu = \kappa^2 T_{\mu\nu} \label{eq:met1s}
\end{equation}
where $T_{\mu\nu}$ is the energy-momentum tensor of the matter. If $f_Q=$constant, taking the trace in this equation one gets an algebraic equation implying that $R=R(T)$, with $T$ the trace of $T_{\mu\nu}$. Because in Palatini $f(R)$ theories the modifications with respect to GR lie in a number of $T$-dependent terms on the right-hand-side of the field equations \cite{or11a}, it follows that for traceless matter-energy sources the dynamics in Palatini theories is exactly the same as in the case of GR ($+$ a cosmological constant term, depending on the Lagrangian chosen). However, nonlinear electrodynamics (NED) are able to yield $T \neq 0$ and thus provide modified dynamics. In such a case the connection equation (\ref{eq:con3}) may be solved by defining a rank-two tensor $h_{\mu\nu}=f_R g_{\mu\nu}$ such that $\nabla_{\mu}(\sqrt{-h} h^{\mu\nu})=0$ and thus $\Gamma_{\alpha\beta}^{\gamma}$ becomes the Levi-Civita connection of $h^{\mu\nu}$. In terms of $h_{\mu\nu}$, the field equations of pure $f(R)$ theories read

\begin{equation}
{R_\mu}^\nu(h)=\frac{1}{f_R^2}\left(\kappa^2 {T_\mu}^\nu+\frac{f}{2}\delta_\mu^\nu \right) \label{eq:met-varRQ4}
\end{equation}
Dealing with this equation for a given gravity and matter Lagrangians, and putting the result in terms of the physical metric $g_{\mu\nu}$, provides a full solution. In \cite{or11a} the particular case of $f(R)=R \pm R^2/R_P$ ($R_P \equiv $ Planck  curvature) coupled to Born-Infeld electrodynamics \cite{BI}  was analyzed, finding the existence of black holes with up to three horizons, and a singularity of minimum divergence $\sim 1/r^2$, which is milder than in GR ($\sim 1/r^4$ for Schwarzschild and $\sim 1/r^8$ for Reissner-Nordstr\"om).

This procedure also works for the general $f(R,Q)$ case. Now there is deviance from GR even for traceless matter-energy sources. The connection equation (\ref{eq:con3}) leads to the following relation between the auxiliary $h_{\mu\nu}$ and physical $g_{\mu\nu}$ metrics

\begin{equation} \label{eq:h-g}
h^{\mu\nu}=\frac{g^{\mu\alpha}{\Sigma_{\alpha}}^\nu}{\sqrt{\det \hat{\Sigma}}} \ , \quad
h_{\mu\nu}=(\sqrt{\det \hat{\Sigma}}){\Sigma_{\mu}}^{\alpha}g_{\alpha\nu} \ .
\end{equation}
where ${\Sigma_{\mu}}^{\nu}=(f_R\delta_{\mu}^{\nu}+2f_Q{P_{\mu}}^{\nu})$ with ${P_\mu}^\nu\equiv R_{\mu\alpha}g^{\alpha\nu}$. In terms of $h_{\mu\nu}$ the field equations read as (\ref{eq:met-varRQ4}) with the replacement $f_R^2 \rightarrow \sqrt{\det(\hat{\Sigma})}$. When $T_{\mu\nu}$  represents a monopolar Maxwell field, a generalization of the Reissner-Nordstr\"om solution is obtained \cite{or12a}. Taking the Lagrangian density $f(R,Q)=R+l_P^2 (R^2 + b Q)$ where $l_P \equiv$ Planck's length and $b$ a constant, solving (\ref{eq:met-varRQ4}) and writing the final result for a static, spherically symmetric metric $ds^2=g_{tt}c^2dt^2+g_{rr}dr^2+r^2d\Omega^2$ leads to
\begin{equation}\label{eq:g}
g_{tt}=-\frac{A(z)}{\sigma_+} \ , \ g_{rr}=\frac{\sigma_+}{\sigma_-A(z)}  \ , \ A(z)=1-\frac{\left[1+\delta_1 G(z)\right]}{\delta_2 z \sigma_-^{1/2}} \ ,
\end{equation}
where $z\equiv r/r_c$, $r_c\equiv \sqrt{r_q l_P}$, $\sigma_{\pm}=1 \pm 1/z^4$, $r_q^2=\kappa^2 q^2/4\pi$ and the parameters
\begin{equation}\label{eq:d1d2}
\delta_1=\frac{1}{2r_S}\sqrt{r_q^3/l_P} \ , \
\delta_2= \frac{\sqrt{r_q l_P} }{r_S} \ .
\end{equation}
where $r_S=2GM/c^2$ and the function $\frac{dG}{dz}=\frac{z^4+1}{z^4\sqrt{z^4-1}}$. These equations define black holes that recover their GR counterparts for $r\gg l_P$ but which undergo relevant non-perturbative changes near $z=1$. Indeed when $\delta_1=\delta_1^* \simeq 0.5720$ an expansion of the curvature invariants about $z=1$ reveals that the spacetime is nonsingular. Moreover, the metric can be extended beyond $z=1$ revealing a wormhole structure whose properties are currently under investigation. 
When $r_q<2l_P$ the event horizon of these objects disappears, making them stable against Hawking decay and with a mass spectrum $ M\approx 1.23605 \left(N_q/N_q^c\right)^{3/2} m_P$, where $N_q<N_q^c\simeq 16.55$ is the number of charges and $m_P$ the Planck mass.

The Palatini approach also allows to study the black hole formation process from a null fluid \cite{or12c} in exact analytical form. Consider an ingoing stream of neutral particles with an energy-momentum tensor $T_{\mu\nu}=\rho_{in} l_{\mu}l_{\nu}$, where $\rho_{in}$ is the energy density of the stream and $l_\mu$ a null radial vector, $l_\mu l^\mu=0$. In this case, the matrix ${\Sigma_\mu}^\nu$ relating the metrics $h_{\mu\nu}$ and $g_{\mu\nu}$ is given by ${\Sigma_\mu}^\nu= \delta_{\mu}^{\nu} + 2l_P^2 \kappa^2 \rho_{in} l_{\mu}l^{\nu}$. Considering a Vaidya-type metric $ds^2=-Be^{2\Psi}dv^2+ 2e^{\Psi}dv dr+r^2 d\Omega^2$ the field equations lead to $\Psi=0$ and

\begin{equation}
B=1-\frac{2\int^{v}L(v')dv'}{r}-\frac{4L(v)}{\rho_P r^2} \label{R-N}
\end{equation}
where $\rho_P=\frac{c^2}{l_P^2 G} \sim 10^{96} kg/m^3$ is Planck's density and $L(v)=\kappa^2 r^2 \rho_{in}/2$ is the luminosity function. Eq.(\ref{R-N}) describes a Reissner-Nordstr\"om solution with a {\it wrong-sign} charge term, that interpolates between two Schwarzschild solutions of masses $M$ and $M'$ through a transient (charge-term) contribution to the mass function.

\section{Concluding remarks}

Palatini theories containing $R$ and $R_{\mu\nu}R^{\mu\nu}$ terms have a rich structure in terms of black hole solutions. In these theories the connection can be algebraically determined and the field equations cast in terms of the metric $h_{\mu\nu}$ associated to a Levi-Civita connection. Transforming back to the physical metric $g_{\mu\nu}$ provides a full solution to a given problem. More results on this line will be presented elsewhere.

Work supported by the Spanish grant FIS2011-29813-C02-02 and the JAE-doc program of the Spanish Research Council (CSIC).

\end{document}